\PassOptionsToPackage{unicode}{hyperref}
\PassOptionsToPackage{hyphens}{url}
\documentclass[12pt]{article}

\usepackage{amsfonts}

\usepackage{amssymb}
\usepackage[utf8]{inputenc}
\usepackage[T1]{fontenc}
\usepackage{times}
\usepackage{lmodern}
\usepackage{latexsym}
\usepackage{amsmath}
\usepackage{amssymb}
\usepackage{tikz}
\usepackage{verbatim}
\usepackage{graphicx}
\usepackage{fancyvrb}

\usepackage{multirow}%
\usepackage{booktabs}%

\usepackage{amscd}
\usepackage{amstext}
\usepackage{amsmath}
\usepackage{caption}
\usepackage{subcaption}
\usepackage{pgfgantt}
\usepackage{multirow}

\usepackage{algorithm}%
\usepackage{algorithmicx}%
\usepackage{algpseudocode}
\usepackage{float} 

\usepackage{amssymb,latexsym}
\usepackage{epsfig}

\DeclareMathOperator{\Tr}{Tr}
\DeclareMathOperator{\Diag}{Diag}

\usepackage{amsmath,amssymb}
\usepackage{xcolor}
\usepackage[margin=1in]{geometry}
\usepackage{color}
\usepackage{fancyvrb}

\DefineVerbatimEnvironment{Highlighting}{Verbatim}{commandchars=\\\{\}}
\usepackage{framed}
\definecolor{shadecolor}{RGB}{248,248,248}

\usepackage{graphicx}

\begin{document}

\title{Shrinkage Estimators for Beta Regression Models}
\author{
	\sc Luis Firinguetti 
	\\
	\small Departamento de Estad\'istica\\
	\small Universidad del B\'{\i}o B\'{\i}o \\
	\small Avda. Collao 1202, Concepci\'on, Chile\\
	\small E-mail:~\texttt{lfiringu@ubiobio.cl}
	\and
	\sc Manuel González-Navarrete\\
	\small Departamento de Matemática y Estad\'i{}stica\\
	\small Universidad de La Frontera \\
	\small Francisco Salazar 1145, Temuco, Chile\\
	\small E-mail:~\texttt{manuel.gonzaleznavarrete@ufrontera.cl}
	\and
	\sc Romer Machaca-Aguilar\\
	\small Dirección de Estadísticas e Indicadores Económicos y Sociales\\
	\small Instituto Nacional de Estadística \\
	\small Avenida José Carrasco 1391, La Paz, Bolivia\\
	\small E-mail:~\texttt{romermachaca8@gmail.com}
	}
\date{}

\maketitle

\begin{abstract}
The beta regression model is a useful framework to model response variables that are rates or proportions, that is to say, response variables which are continuous and restricted to the interval (0,1). As with any other regression model, parameter estimates may be affected by collinearity or even perfect collinearity among the explanatory variables. To handle these situations shrinkage estimators are proposed. In particular we develop ridge regression and LASSO estimators from a penalized likelihood perspective with a logit link function. The properties of the resulting estimators are evaluated through a simulation study and a real data application.
\end{abstract}

\section{Introduction}\label{sec1}

Regression modeling is a versatile and powerful tool for statistical analysis that seeks to explain the relationship between a dependent or response variable and one or more independent or explanatory variables. In its most common form it is assumed that the dependent variable can take any value on the real line, while the explanatory variables can be of any type. In this context it is possible, and perhaps plausible, to assume the dependent variable is normally distributed. In practice however, there are many applications in disciplines such as public health, biology, ecology, economics and marketing, where the response variable, being a proportion, a rate or a probability, is restricted to lie in the interval (0, 1). It is in this context that Ferrari \& Cribari-Neto \cite{ferrari} introduce the beta regression model, which assumes a response variable with a beta distribution. This distribution is extremely flexible, allowing a wide variety of shapes of data distributions to be modeled, from uniform to highly skewed distributions. Thus, beta regression takes advantage of this flexibility to better fit proportional data. This methodology clearly differs from other types of regression due to the particular characteristics of the variables being studied, which cannot take values outside the range (0, 1), making traditional regression methods not appropriate.

Besides, in the classical linear regression model (CLRM) multicollinearity is a common problem, resulting in ordinary least squares (OLS) and maximum likelihood (ML) estimators with large variances. To handle estimation in a high dimensional setting and to deal with the problem of collinearity, shrinkage estimators, that may be found as a solution to a penalized least squares regression problem, have often been proposed. They are expected to produce significant reductions in the variance of parameter estimates, although at the expense of introducing some bias. As a consequence, they may have a lower mean square error than the OLS or ML estimators. Among these alternative estimators we have: the garotte introduced by Breiman \cite{garotte}; the ridge regression (RR) estimator developed by Hoerl \& Kennard \cite{hoerl}; the least absolute shrinkage and selection operator (LASSO) estimator proposed by Tibshirani \cite{Tibshirani}; the elastic net proposed by Zou \& Hastie \cite{Hastie}. It is true however that the problem of multicollinearity is not exclusive of the CLRM as it may also be present in the context of a generalized linear model (GLM). In the event that multicollinearity is present, Schaefer et. al  \cite{notasrid} and Le Cessie \& Van Houwelingen \cite{logistico}, propose using ridge estimators in the logistic regression model. Also Friedman et. al. \cite{friedman} applied the LASSO method to cases where the dependent variable is distributed according to a probability distribution function belonging to the exponential family. These estimators are expected to be more efficient than the classical ML estimators.

In this paper we focus on a regression model with a beta distributed response variable, with mean $\mu$ and a precision parameter $\phi$. As declared earlier, this approach was introduced by Ferrari \& Cribari-Neto \cite{ferrari} , who carried out the estimation of the parameters, by maximizing the log-likelihood using the Newton Raphson’s iterative method. Again, we argue as before, that the precision of the ML estimator may well be affected by collinearity. To tackle this problem Qasim et al. \cite{betarid} as well as Abonazel \& Taha \cite{taha} applied the ridge method in the beta regression model (see also \cite{reza}). In their approach the RR estimators are obtained by minimizing the length of the coefficient estimator subject to a quadratic constraint. In this paper we approach the estimation problem from a different perspective: the RR estimator is obtained by maximizing  the log-likelihood function with a suitable penalization. In addition to RR, we also derive the LASSO estimator by maximizing a penalized log-likelihood.

The rest of the paper is organized as follows: in the next section we introduce the beta regression model; in Section 3 we develop our RR and LASSO estimators; in Section 4 we compare our estimators with the ordinary ML estimator through simulations; in Section 5 we provide an application of the proposed estimators, and finally in Section 6 we conclude with some final remarks.

\section{ Beta regression model}

\subsection{Beta distribution}

Let us start by recalling the beta distribution. A random variable $Y$ is said to have a beta distribution with parameters $a>0$ and $b>0$, ($Y \sim beta(a, b)$) if its density function is given by:

\begin{equation} \label{distr_beta}
	f(y;a,b)=\dfrac{\Gamma(a+b)}{\Gamma(a)\Gamma(b)}y^{a-1}(1-y)^{b-1}=\dfrac{1}{B(a,b)}y^{a-1}(1-y)^{b-1},
\end{equation}
where $\Gamma$ is the gamma function and $B(a,b)$ is the beta function.

For practical purposes, the following reparametrization is made: $\mu=\dfrac{a}{a+b}$ and $\phi= a+b$, hence $a=\mu\phi$ and $b=(1-\mu)\phi$. The mean and variance of $Y$ are $E[Y]=\mu$ and $V[Y]=\dfrac{\mu(1-\mu)}{1+\phi}$, for $\phi>0$ is the precision parameter. The density function written with this new reparameterization is:
\begin{equation}\label{d_beta}
	f(y,\mu, \phi)=\dfrac{\Gamma(\phi)}{\Gamma(\mu\phi)\Gamma((1-\mu)\phi)} y^{\mu\phi-1}(1-y)^{(1-\mu)\phi-1},0<y<1.
\end{equation}

\subsection{The beta regression model }

In this section we consider the beta regression model proposed by Ferrari and Cribari-Neto \cite{ferrari}. Let $y_1, y_2, \ldots , y_n$ be $n$ observations on the dependent variable which are assumed to be independent random variables following the beta density function defined in the equation (\ref{d_beta}) with  mean $\mu$ and precision parameter $\phi$. Let also  $x_i^{t} = \left(1 \quad x_{i1} \quad x_{i2} \quad \ldots \quad x_{ip} \right)$, be a vector of observations on $p$ explanatory variables. Although not strictly necessary, we will assume that $p<n$. The design matrix is then given by:

$$
X=\left(\begin{matrix}
	 x_{1}^{t} \\ x_{2}^{t} \\ \vdots \\ x_{n}^{t}
\end{matrix}\right) = \left(\begin{matrix}
	1 & x_{11} & x_{12} & \cdots & x_{1p} \\ 1 & x_{21} & x_{22} & \cdots & x_{2p} \\
	\vdots & \vdots & \vdots & \ddots & \vdots \\ 1 & x_{n1} & x_{n2} & \cdots & x_{np}
\end{matrix}\right),
$$

\noindent $y$ and $X$ will be related through a link function $g(\mu_t)=x_t^{t}\beta$ with $\beta = \left(\beta_0 \quad \beta_1 \quad \beta_2 \quad \ldots \quad \beta_p \right)$ the parameters vector. In this paper we will use the logit function which is monotonic and at least twice differentiable $g(\mu_t) =\log\left(\frac{\mu_t}{1-\mu_t}\right)$ and solving for $\mu_t$ we have $\mu_t = \frac{\exp{(x_t^{t}\beta)}}{1+\exp{(x_t^{t}\beta)}}$.

Once the structure of the model has been built, the parameter vector $\beta$ is estimated using maximum likelihood. However, this estimator also depends on the precision parameter $\phi$. Therefore, it is necessary to estimate the latter as well.

\subsection{The maximum likelihood estimator}

The log-likelihood function is 
\begin{equation}\label{log_mod}
	l(\beta,\phi)=\sum \limits_{t=1}^n l_t(\mu_t,\phi),
\end{equation}
where
\begin{eqnarray}\label{logmod1}
	l_t(\mu_t,\phi)&=&\log \Gamma(\phi)-\log \Gamma(\mu_t\phi)-\log \Gamma((1-\mu_t)\phi)+(\mu_t\phi-1)\log y_t+ \label{log_beta}\\
	&& [(1-\mu_t)\phi-1]\log(1-y_t).\nonumber
\end{eqnarray}
Due to its non linearity the Fisher scoring method is used to obtain a numerical solution:
$$
\hat{\theta}^{(k+1)}=\hat{\theta}^{(k)}+\mathfrak{J}^{-1}(\hat{\theta}^{(k)}) \left. \frac{\partial l(\theta)}{\partial \theta}\right|_{\theta=\hat{\theta}^{(k)}} \mbox{, } k=0,1,\ldots,
$$
where
$ \hat{\theta}^{(k)}=(\hat{\beta}^{(k)} \quad \hat{\phi}^{(k)})^t $
is the  $k$-th iteration and $\mathfrak{J}^{-1}$ is the inverse of the Fisher information matrix. If $\phi$ is assumed to be known then the estimator of $\beta$ is 
\begin{equation}\label{est_mle}
	\hat{\beta}_{ML}= (X^{t}WX)^{-1}X^{t}W z,
\end{equation}
\noindent where $W=\Diag(w_1,\ldots,w_n)$, $w_t=\phi\left\{\psi^\prime(\mu_t\phi)+\psi^\prime((1-\mu_t)\phi)\right\} \dfrac{1}{(g^\prime(\mu_t))^2}$ with $\psi^\prime()$ the trigamma function and $z=(g(y_1),\ldots, g(y_n))^{t}$.


\section{Shrinkage estimators}

As argued before, shrinkage methods have been proposed as a solution to two problems that are fairly frequent in regression models. The first is related to the existence of collinearity among the explanatory variables, causing the variance to be large and hence producing estimators which are not very reliable. The second problem is caused by high dimensional $X$ matrices, where the number of variables exceed the number of observations. This is in fact a case of perfect multicollinearity.

\subsection{Ridge regression}
We approach the estimation problem from the perspective of a penalized likelihood. Then, consider maximizing the log-likelihood $l(\beta)$ penalized by the term $k\|\beta\|^2$, that is,
$$
l^{k}(\beta)=l(\beta)-k\|\beta\|_2,
$$
where $\|\beta\|_2 = (\sum\beta_j^2)^{\frac{1}{2}}$ is the $L^2$ norm of the parameter vector $\beta$ and $k$ is the shrinkage parameter. We calculate the second derivative of the penalized log-likelihood and applying Newton Raphson gives the RR estimator in the beta regression model (see appendix A)

\begin{equation}\label{betrig}
	\hat{\beta}_{ridge} =\left(X^{t}WX+kI\right)^{-1}X^{t}WX\hat{\beta}_{ML}.
\end{equation}
\noindent where $I$ is the identity matrix of dimension $p+1$ and assumes that $\phi$ is known.\\


We can prove that the beta ridge estimator performs better than the beta estimator without penalties, by calculating their biases, variances and subsequently comparing their mean squared errors (MSE). For more details review \cite{betarid} and \cite{taha}.\\

In direct reference to  Hoerl \& Kennard \cite{hoerl}, Qasim, Mansson \& Kibria \cite{betarid} and independently Abonazel \& Taha \cite{taha} propose a version of the RR estimator for the beta regression model which coincides with the estimator presented in (\ref{betrig}), with the difference that these authors arrive at this result by minimizing the squared error, and in this article the likelihood is maximized.

\subsection{The LASSO estimator }

To obtain the LASSO estimator in the beta regression model, we adopt the strategy followed by Noah et. al. \cite{noah} and minimize a modified negative log-likelihood  penalized by the $L^1$ norm, $\|\beta\|_1 = \sum_{j}|\beta_j|$. Thus, the LASSO estimator is obtained by minimizing the following objective function:

\begin{equation}\label{decent}
	\min\left(-\frac{1}{n}l(\beta)+k\|\beta\|_1\right).
\end{equation}

To proceed, we make a second order Taylor series expansion of the log-likelihood $l(\beta)$ (see appendix B) with $\tilde{\beta}$ and $\tilde{\eta}$ some estimators of $\beta$ and $\eta = X\beta$ respectively.

\begin{eqnarray}
	l(\beta)&\simeq & l(\tilde{\beta})+(\beta-\tilde{\beta})^{t} l^\prime(\tilde{\beta})+\frac{(\beta-\tilde{\beta})^{t} l^{\prime\prime}(\tilde{\beta})(\beta-\tilde{\beta})}{2}, \\ \label{taylor}
	l(\beta) & \simeq & \frac{1}{2}\left[z(\tilde{\eta}) - \eta \right]^{t} l^{\prime\prime}(\tilde{\eta})\left[z(\tilde{\eta}) - \eta \right] + C(\tilde{\eta}),\label{taylor2}
\end{eqnarray}
with $z(\tilde{\eta}) = \tilde{\eta} -l^{\prime\prime}(\tilde{\eta})^{-1}l^{\prime}(\tilde{\eta})$ and $C(\tilde{\eta})=l(\tilde{\beta}) -\frac{1}{2} l^\prime(\tilde{\eta})^{t}l^{\prime\prime}(\tilde{\eta})^{-1}l^\prime(\tilde{\eta})$.\\

Then, using the coordinate descent method (method used in \cite{friedman} and \cite{noah}) we find the LASSO estimate of $\beta_j$ assuming the remaining $\beta_i$ $\forall i\not=j$ and $\phi$ are known.
If we denote by $X_{-j}$ the $X$ matrix with the $j$-th column removed and by $\tilde{\beta}_{-j}$ the corresponding vector of estimated coefficients. Then, the log-likelihood may be written as follows:
\begin{equation}
	l(\beta) \simeq \frac{1}{2}\left[z(\tilde{\eta}) - X_{-j}\tilde{\beta}_{-j} - X_{j}\beta_{j} \right]^{t} l^{\prime\prime}(\tilde{\eta})\left[z(\tilde{\eta}) - X_{-j}\tilde{\beta}_{-j} - X_{j}\beta_{j} \right] + C(\tilde{\eta}),    
\end{equation}
and then,
\begin{equation}
	\dfrac{\partial l(\beta)}{\partial \beta_j} \simeq -X_j^{t}l^{\prime\prime}(\tilde{\eta})\left[z(\tilde{\eta}) - X_{-j}\tilde{\beta}_{-j} - X_{j}\beta_{j} \right].
    \end{equation}
We obtain a LASSO-type estimator minimizing the following expression: 
\begin{equation}
	\hat{\beta}_{lasso} = argmin\left\{-\frac{1}{n}l(\beta)+k|\beta_j|+k \sum\limits_{i\not=j}|\beta_i|\right\}.		
    \end{equation}
Differentiating this expression and equating to 0, gives:
\begin{equation}
	\frac{1}{n} X_j^{t} l^{\prime\prime}(\tilde{\eta}) X_{j}\beta_{j}= \frac{1}{n} X_j^{t} l^{\prime\prime}(\tilde{\eta})\left[z(\tilde{\eta}) - X_{-j}\tilde{\beta}_{-j}\right] +  k \cdot \text{sign}(\beta_j).
    \end{equation}
where
$$
\text{sign}(x) = \left\{ \begin{array}{ll}
	-1      & \mbox{ si } x<0, \\
	\left[-1,1\right] & \mbox{ si } x=0, \\
	1     & \mbox{ si } x>0,
\end{array} \right.
$$
and changing the second order derivatives with respect to $x\tilde{\beta}$ for its expected value, we have:
\begin{equation}
	\beta_{j} \simeq \dfrac{\mathbb{S}_{k} \left(\frac{1}{n}X_j^{t}\Omega(\tilde{\eta})\left[z(\tilde{\eta}) - X_{-j}\tilde{\beta}_{-j}\right]\right)}{\frac{1}{n} X_j^{t} \Omega(\tilde{\eta})X_{j}}, \label{lasso}
    \end{equation}
with $z(\tilde{\eta})=\tilde{\eta} -\Omega(\tilde{\eta})^{-1}l^{\prime}(\tilde{\eta})$, $\Omega(\tilde{\eta}) = E(l^{\prime\prime}(\tilde{\eta}))$ and
the soft-thresholding operator
$$
\mathbb{S}_k(x)=\left\{
\begin{array}{ll}
	x+k & \mbox{ si } x< - k, \\
	0 & \mbox{ si } |x| \leq k, \\
	x-k & \mbox{ si } x > k.
\end{array}
\right. 
$$
Thus, replacing the vector of first derivatives  (\ref{score_xb}) and the expected value of the matrix of second order derivatives we obtain the estimator for $\beta_j$ coordinate:
\begin{equation}\label{est_lasso}	
	\hat{\beta}_{lasso(j)}  \simeq \dfrac{\mathbb{S}_{k} \left( -\frac{\phi}{n} X_j^{t} W \left[z(\tilde{\eta}) - X_{-j}\tilde{\beta}_{-j}\right]\right)}{- \frac{\phi}{n}X_j^{t} W X_{j}},
\end{equation}
where $z(\tilde{\eta}) =(X\tilde{\beta})^{t} -W^{-1}T (y^\ast-\mu^\ast)$. The initial value of  $\hat{\beta}_{lasso}$ is the maximum likelihood estimator $\hat{\beta}_{ML}$.

We also use the coordinate descent method to estimate the precision parameter $\phi$. This is done by maintaining fix the estimated $\tilde{\beta}_j$ while estimating $\phi$. Summarizing, the proposed LASSO estimator is obtained through the algorithm 1.

\begin{algorithm}
\caption{Calculate $\hat{\beta}_{lasso}$}\label{algo1}
\begin{algorithmic}[2]
\Require Initial value $\hat{\beta}_{lasso}^{(0)} = \hat{\beta}$
\While{convergence}
        \ForAll{$j=0, 1, 2, \ldots, p$}
            \State Compute $W(\hat{\beta}_{lasso(j)}^{(s-1)})$ and $z(X\hat{\beta}_{lasso(j)}^{(s-1)})$
            \State Find $\hat{\beta}_{lasso(j)}^{(s)}$ from equation (\ref{est_lasso})
            \State Compute $\hat{\beta}_{lasso}^{(s)}-\hat{\beta}_{lasso}^{(s-1)}$
        \EndFor
    \State Minimize $-\frac{1}{n}l(\hat{\beta}_{lasso(j)}^{(s)})+k|\hat{\beta}_{lasso(j)}^{(s)}|$ with respect to $\hat{\phi}_{lasso}^{(s)}$
\EndWhile
\end{algorithmic}
\end{algorithm}

\section{A simulation study}
A simulation is carried out to analyse the performance of the proposed estimators.

\subsection{Simulation design}
In this study we consider models with the following specifications of the elements defining the beta regression model:
 \begin{itemize}
 	\item 
 	For the precision parameter $\phi$, we specify two values: $\phi \in \{1,5\}$.
 	\item We specify models with $p=5$ and $p=7$ variables, including a constant term. Then,	$\underline{\beta}=\left(\beta_0, \beta_1, \ldots, \beta_p\right)^t$, with $\beta_0=0$, $\sum_{j=1}^{p}\beta_j^2 = 1$. To evaluate the ability of the lasso to select variables, the coefficients of two last variables are set to zero. All non-zero $\beta_j$ are assumed to be equal.
  
 
 \item
 	 The explanatory variables were generated from the standard normal distribution.
 \item To achieve different degrees and patterns of multicollinearity among the explanatory variables, we specified different values of $\rho$ ($\rho\in \{0.1, 0.2, \ldots, 0.9\}$).
 \item 
 	To generate the observations in the dependent variable, $y_i$,  with a beta distribution with a mean vector $\mu =(\mu_1  \ldots  \mu_n)$, with
$$ \mu_{i} = \dfrac{\exp(x_i^{t}\beta)}{1+\exp(x_i^{t}\beta)}, \quad i=1,2, \ldots n.
$$
That is,  $y_i\sim beta(\alpha=\mu_i\phi, \beta=(1-\mu_i)\phi))$.
\end{itemize}

Each model specification was replicated 2,000 times. The following estimators are evaluated:
\begin{itemize}
	\item
	The maximum likelihood estimator in \eqref{est_mle}.
	\item 	The ridge regression estimator in \eqref{betrig}, using four criteria to select the shrinkage parameter, namely: $k_{HK} = \frac{1}{\hat{\phi}\hat{\gamma}^2_{\max}}$, from \cite{hoerl}; $k_{med} = Median \left(\sqrt{\frac{1}{\hat{\phi}\hat{\gamma}^2_{0}}},\,\sqrt{\frac{1}{\hat{\phi}\hat{\gamma}^2_{1}}},\,\sqrt{\frac{1}{\hat{\phi}\hat{\gamma}^2_{2}}},\,\dots\,,\,\sqrt{\frac{1}{\hat{\phi}\hat{\gamma}^2_{p}}}\right)$, from \cite{kibria}; $k_{\max} = \dfrac{\lambda_{\max}}{\hat{\phi}\hat{\gamma}^2_{\max}}$ and $k_{\min} = \dfrac{\lambda_{\min}}{\hat{\phi}\hat{\gamma}^2_{\min}}$, from \cite{taha}.
	\item The LASSO estimator in Algorithm 1, using cross-validation to select the shrinkage parameter.
\end{itemize}
We remark that the shrinkage estimators are calculated using standardized variables and then converted back to the original variables.\\

To compare these estimators we make use of mean square error ($MSE$), which for any estimator $\hat{\beta}$ is calculated as $MSE(\hat{\beta}_{j})=\sum_{i=1}^{2000}(\hat{\beta}_{ji}-\beta_{j})^{2}/2000$, with $j=0,1,\ldots,p$ and total mean square error ($TMSE$), which for an estimator $\hat{\beta}$ is calculated by $TMSE(\hat{\beta})=\sum_{j=0}^{p}MSE(\hat{\beta}_{j})$.

	\subsection{Simulation results}
We develop three different kinds of comparison. First, we analyse the performance of RR with the four criteria to select the shrinkage parameter. Then, we compare the MLE, RR and LASSO with different levels of collineality. Finally, we study the ability of our LASSO estimator in the variable selection context.

\subsubsection{Comparison of ridge regression estimators with different shrinkage parameter}
We start comparing the performance of RR estimators by using different criteria to select the shrinkage parameter.

\begin{table}[h]
\caption{TMSE comparison of maximum likelihood and ridge regression estimators for $n=30,50,100,200$; $\phi=1$; $p=5$ and three levels of multicollinearity.}\label{c01}%
		\begin{tabular}{ l  l  r  r  r  r  r }
			\hline 
			\multirow{2}{*}{Multicollinearity} & \multirow{2}{*}{n} & MLE & \multicolumn{4}{c}{Ridge Regression}\\ \cmidrule{4-7}
			& & $k=0$ &  $k_{HK}$ & $k_{med}$ & $k_{max}$ & $k_{min}$ \\ \hline
			\multirow{4}{*}{Low ($\rho=0.4$)}& 30 & 0.3441 & 0.3460 & 0.3337 & 0.7116 & 0.4961 \\
			& 50 & 0.3048 & 0.3109 & 0.2997 & 0.7863 & 0.4996 \\
			& 100 & 0.2052 & 0.2291 & 0.2190 & 0.8368 & 0.5429 \\
			& 200 & 0.1905 & 0.2112 & 0.2014 & 0.8829 & 0.5526 \\ \hline
			\multirow{4}{*}{Moderate ($\rho=0.7$)}& 30 & 0.4858 & 0.3868 & 0.3477 & 0.6906 & 0.3952 \\
			& 50 & 0.4823 & 0.4036 & 0.3833 & 0.8232 & 0.4331 \\
			& 100 & 0.2812 & 0.2727 & 0.2667 & 0.8472 & 0.3553 \\
			& 200 & 0.2603 & 0.2629 & 0.2582 & 0.8958 & 0.3941 \\ \hline
			\multirow{4}{*}{High ($\rho=0.9$)}& 30 & 1.1581 & 0.6361 & 0.4472 & 0.6236 & 0.7266 \\
			& 50 & 0.9796 & 0.5220 & 0.4529 & 0.7772 & 0.5797 \\
			& 100 & 0.4752 & 0.3483 & 0.3250 & 0.8067 & 0.3313 \\
			& 200 & 0.3653 & 0.3116 & 0.3023 & 0.8653 & 0.3045 \\ \hline
		\end{tabular}
\end{table}

\begin{table}[h]
		\caption{TMSE comparison of maximum likelihood and ridge regression estimators for $n=30,50,100,200$; $\phi=5$; $p=5$ and three levels of multicollinearity.}\label{c02}%
		\begin{tabular}{ l  l  r  r  r  r  r  r }
			\hline 
			\multirow{2}{*}{Multicollinearity} & \multirow{2}{*}{n} & MLE & \multicolumn{4}{c}{Ridge Regression}\\ \cmidrule{4-7}
			& & $k=0$ &  $k_{HK}$ & $k_{med}$ & $k_{max}$ & $k_{min}$ \\ \hline
			\multirow{4}{*}{Low ($\rho=0.4$)} & 30 & 0.1588 & 0.1541 & 0.1456 & 0.1724 & 0.1519 \\
			& 50 & 0.1174 & 0.1159 & 0.1138 & 0.1676 & 0.1167 \\
			& 100 & 0.0571 & 0.0580 & 0.0584 & 0.1516 & 0.0807 \\
			& 200 & 0.0409 & 0.0418 & 0.0421 & 0.1654 & 0.0669 \\ \hline
			\multirow{4}{*}{Moderate ($\rho=0.7$)} & 30 & 0.2701 & 0.2450 & 0.2083 & 0.2083 & 0.2405 \\
			& 50 & 0.2258 & 0.2111 & 0.1962 & 0.2482 & 0.2013 \\
			& 100 & 0.1016 & 0.0993 & 0.0975 & 0.1989 & 0.0968 \\
			& 200 & 0.0757& 0.0757 & 0.0756 & 0.2278 & 0.0793 \\ \hline
			\multirow{4}{*}{High ($\rho=0.9$)} & 30 & 0.7630 & 0.6237 & 0.3989 & 0.3062 & 0.6894 \\
			& 50 & 0.5594 & 0.4438 & 0.3171 & 0.2802 & 0.4734 \\
			& 100 & 0.2357 & 0.2091 & 0.1789 & 0.2141 & 0.2044 \\
			& 200 & 0.1487 & 0.1407 & 0.1319 & 0.2321 & 0.1271 \\ \hline
		\end{tabular}
\end{table}

In particular, in Tables \ref{c01} and \ref{c02} we observe that in low collinearity, the MLE estimator presents better results in the sense of the TMSE. However, if multicollinearity is moderate or high, the RR estimator is better than the MLE. We also see that the larger the value of $n$, the smaller the TMSE of all estimators. The same happens for $\phi$, for larger values of $\phi$ the TMSE tends to be lower. Comparing the different options of the shrinkage parameter $k$, we can see that the values of $k$ proposed by Muniz \& Kibria \cite{taha}, namely $k_{med}$, presents a better performance taking the TMSE as measure of goodness.

\subsubsection{Comparison of maximum likelihood, ridge regression and LASSO estimators.}

\begin{figure}[h!]
	\centering
	\includegraphics[width=0.9\textwidth]{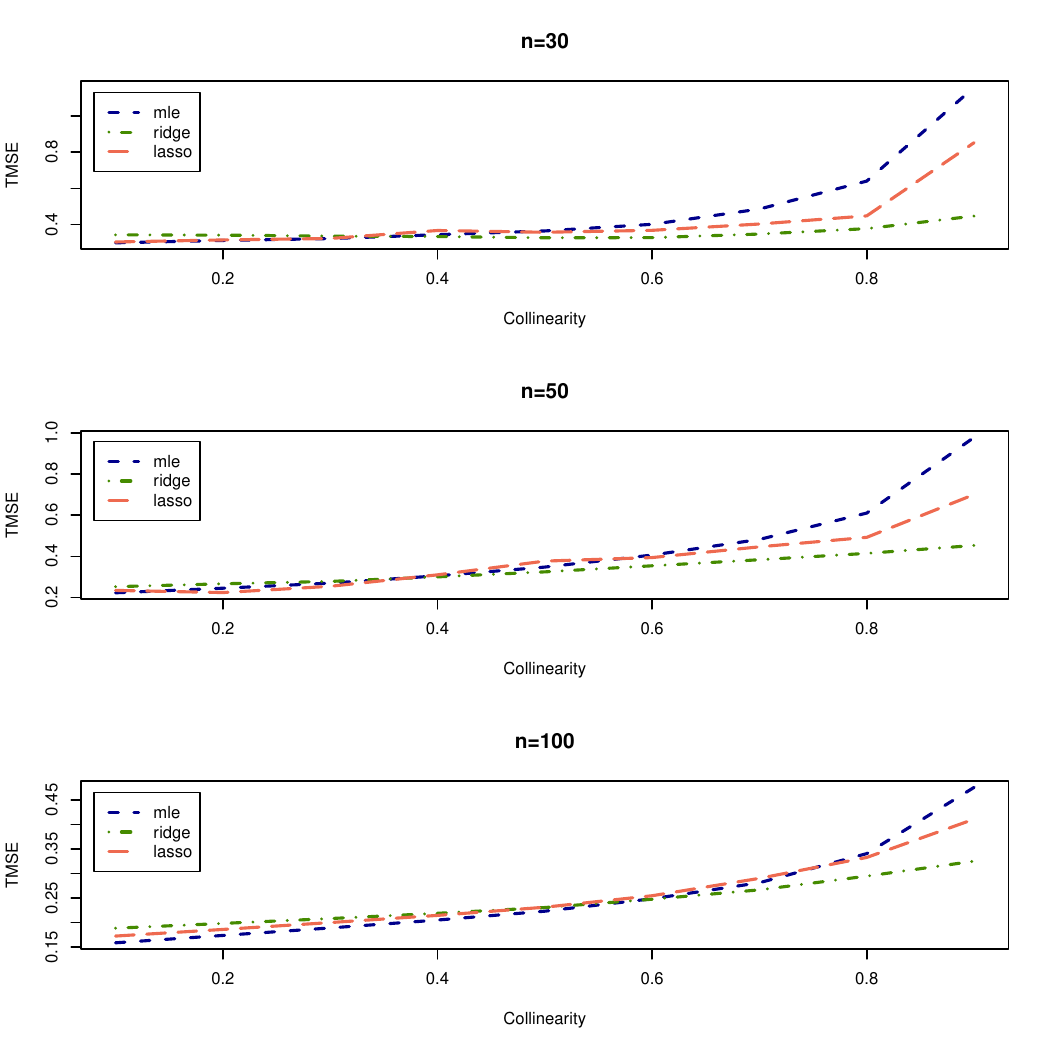}
	\caption{$TMSE$ comparison of maximum likelihood, ridge regression and LASSO estimators for  $n=30, 50, 100$; $\phi=1$; $p=5$ as a function of levels of multicollinearity.}
	\label{graf_phi1_p4}
\end{figure}

\begin{figure}[h!]
	\centering
	\includegraphics[width=0.9\textwidth]{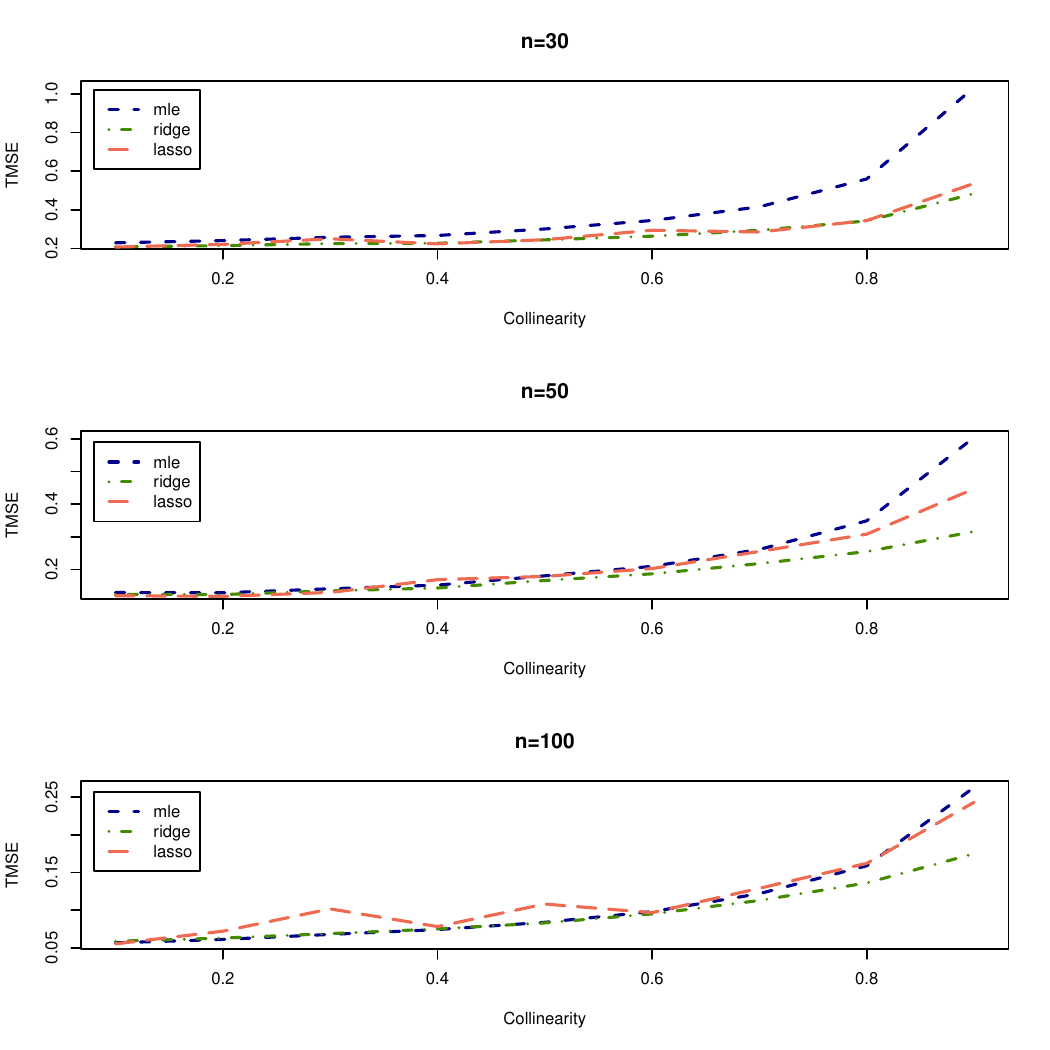}
	\caption{$TMSE$ comparison of maximum likelihood, ridge regression and LASSO estimators for  $n=30, 50, 100$; $\phi=5$; $p=7$ with $\beta_5 = \beta_6 = 0$ as a function of levels of multicollinearity.}
	\label{graf_phi5_p6}
\end{figure}

In this case we consider the RR estimator with  shrinkage parameter $k=k_{med}$. The results may be found in Figures \ref{graf_phi1_p4}-\ref{graf_phi5_p6}. In these figures we compare the TMSE of each estimator.

Note from these figures that, in most cases, RR has a lower $TMSE$ than the other two estimators followed by the LASSO estimator, which in most cases has a lower TMSE than MLE. Another important aspect shown in the tables is the behavior of the $TMSE$ with respect to the sample size; as $n$ increases, the $TMSE$ decreases. However, this feature is not very clear in the LASSO estimator.

We may also confirm a previous finding: as the precision parameter $\phi$ increases the $TMSE$ diminishes for all estimators. With increasing multicollinearity we find that RR outperform LASSO, which in turn outperforms MLE.


\subsubsection{Analysis of variable selection with LASSO.}

The Table \ref{c03} summarizes the ability of LASSO to select explanatory variables. In this case, the number of parameters estimated is 7, but where the last two are zero ($\beta_5 = \beta_6 = 0$). 
\begin{table}[ht]
		\caption{Percentage of times that LASSO estimates $\beta_{5}=0$ and/or $\beta_{6}=0$ with $p=6$, $n=30, 50, 100$, $\phi=5$ as a function of levels of multicollinearity.}\label{c03}%
		\begin{tabular}{ c   r  r  r }
			\hline 
			Multicollinearity & $n=30$ & $n=50$ & $n=100$ \\ \hline
			0.1 & 64.28 & 79.73 & 94.60 \\
			0.2 & 64.95 & 80.28 & 95.55 \\
			0.3 & 66.70 & 81.48 & 96.28 \\
			0.4 & 67.05 & 70.45 & 55.32 \\
			0.5 & 68.82 & 66.28 & 86.08 \\
			0.6 & 70.10 & 42.65 & 85.60 \\
			0.7 & 69.92 & 71.30 & 95.98 \\
			0.8 & 70.90 & 49.20 & 95.15 \\
			0.9 & 71.27 & 55.45 & 94.75 \\
			\hline
		\end{tabular}
\end{table}
It may be observed that  the percentage the lasso estimates $\beta_{5}=\beta_{6}=0$, tends to increase with $n$. Also, the higher is the degree of multicolinerality in the design matrix, the larger tends to be the percentage LASSO estimates $\beta_{5}=\beta_{6}=0$. This is an advantage of LASSO over ridge, which cannot estimate any parameter as zero.

\section{An application}
We now present an application with data that contains 506 census tracts of Boston from the 1970 census, data set comes with the package MASS of the R software. The 12 variables included in the analysis are the following:

\begin{itemize}
\item {\textbf{lstat ($y$).- }} lower status of the population (percent).
\item {\textbf{crim ($x_{1}$).- }} per capita crime rate by town.
\item {\textbf{zn ($x_{2}$).- }} proportion of residential land zoned for lots over $25,000$ sq.ft.
\item {\textbf{indus ($x_{3}$).- }} proportion of non-retail business acres per town.
\item {\textbf{nox ($x_{4}$).- }} nitrogen oxides concentration (parts per 10 million).
\item {\textbf{rm ($x_{5}$).- }} average number of rooms per dwelling.
\item {\textbf{age ($x_{6}$).- }} proportion of owner-occupied units built prior to 1940.
\item {\textbf{dis ($x_{7}$).- }} weighted mean of distances to five Boston employment centres.
\item {\textbf{tax ($x_{8}$).- }} full-value property-tax rate per \$ $10,000$.
\item {\textbf{ptratio ($x_{9}$).- }} pupil-teacher ratio by town.
\item {\textbf{black ($x_{10}$).- }} $1000 (Bk - 0.63)^2$ where $Bk$ is the proportion of blacks by town.
\item {\textbf{medv ($x_{11}$).- }} median value of owner-occupied homes in \$ $1000$s.
\end{itemize}

First of all, we verify how the dependent variable is distributed, the Kolmogorov-Smirnov test is applied to test whether it has a normal or beta distribution. By evaluating the hypothesis test that indicates that the data are normally distributed, we obtain $p_{value}= 2.2e^{-16}$ then, the hypothesis that the data have a normal distribution is rejected.\\

Now, we test the hypothesis that indicates that the data is beta distributed, we have $p_{value}= 0.7763$ indicating that it is accepted that the data have a beta distribution.

In addition, we analyse the multicollinearity of the data. Therefore, we compute the simple correlations among the explanatory variables, shown next:

\begin{figure}[h!]
	\centering
	\includegraphics[width=0.7\textwidth]{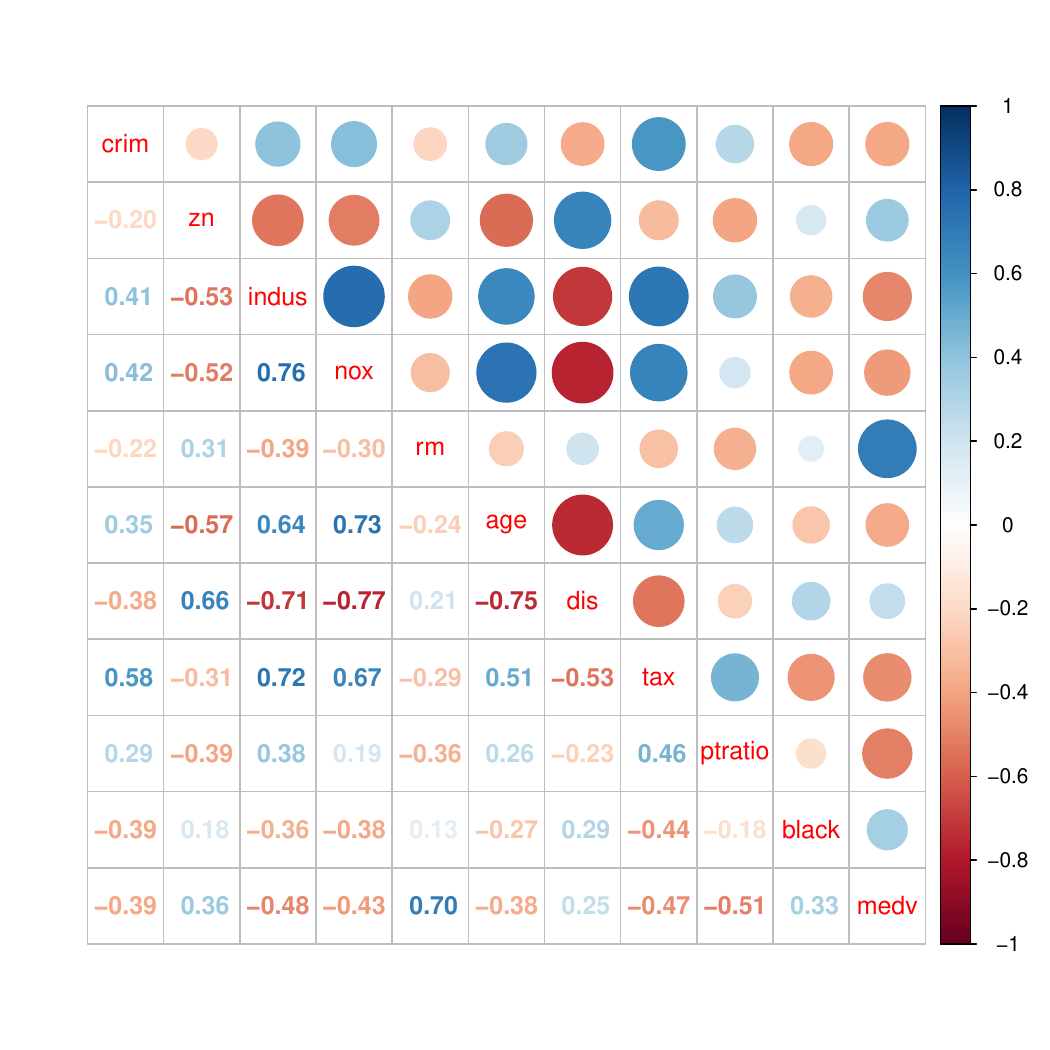}
	\caption{Scatterplot for design matrix}
	\label{graf_corr}
\end{figure}

We can see that the largest correlation is between  $x_4$ and $x_7$ ($-0,77$). However, a better measure of multicollinearity  is the condition number, that is, $h(X) = \sqrt{\dfrac{\lambda_{\max}}{\lambda_{\min}}} = 8198.872,$,
where $\lambda_{\max}$ and $\lambda_{\min}$ are respectively the largest and smallest eigenvalue of  $X^{t}X$. This value is sufficiently large to indicate high multicollinearity.

We now estimate the beta regression model using the MLE, RR and the LASSO estimators. The RR estimator is obtained with $k=k_{\max}=0.1580$ and for the LASSO we use cross validation ($k=k_{cv}=0.04$). The estimation results are given in Table \ref{c_est}.

\begin{table}[h!]
		\caption{Maximum Lihelihood, Ridge Regression and LASSO Estimates}\label{c_est}%
		\begin{tabular}{c l rrr}
			\hline
\multirow{2}{*}{Parameter} &  \multirow{2}{*}{Description} &  \multicolumn{3}{c}{Estimators}\\ \cmidrule{3-5}
&& MLE & RR & LASSO \\ \hline
\multirow{2}{*}{$\beta_0$} & \multirow{2}{*}{Intercept} & 0,21254 & 0,19952 & -0,12524\\
&  & {\footnotesize (0,8224)} & {\footnotesize (0,7488)} & {\footnotesize (-0,3740)}\\ \cmidrule{2-5}
\multirow{2}{*}{$\beta_1$} & \multirow{2}{*}{crim} & -0,00089 & -0,00085 & 0\\
&  & {\footnotesize (0,0002)} & {\footnotesize (-0,0001)} & {\footnotesize -}\\ \cmidrule{2-5}
\multirow{2}{*}{$\beta_2$} & \multirow{2}{*}{zn} & -3,397E-06 & -3,398E-06 & 0\\
&  & {\footnotesize (-4,261E-08)} & {\footnotesize (-4,261E-08)} & {\footnotesize -}\\ \cmidrule{2-5}
\multirow{2}{*}{$\beta_3$} & \multirow{2}{*}{indus} & 0,00315 & 0,00320 & 0,00264\\
&  & {\footnotesize (0,0011)} & {\footnotesize (0,0012)} & {\footnotesize (0,0009)}\\ \cmidrule{2-5}
\multirow{2}{*}{$\beta_4$} & \multirow{2}{*}{nox} & -0,31534 & -0,30561 & -0,03328\\
&  & {\footnotesize (-1,0056)} & {\footnotesize (-0,9628)} & {\footnotesize (-0,0391)}\\ \cmidrule{2-5}
\multirow{2}{*}{$\beta_5$} & \multirow{2}{*}{rm} & -0,20249 & -0,20265 & -0,1984\\
&  & {\footnotesize (-0,5056)} & {\footnotesize (-0,5068)} & {\footnotesize (-0,4897)}\\ \cmidrule{2-5}
\multirow{2}{*}{$\beta_6$} & \multirow{2}{*}{age} & 0,00706 & 0,00703 & 0,00676\\
&  & {\footnotesize (0,0035)} & {\footnotesize (0,0035)} & {\footnotesize (0,0033)}\\ \cmidrule{2-5}
\multirow{2}{*}{$\beta_7$} & \multirow{2}{*}{dis} & -0,02188 & -0,02168 & -0,01486\\
&  & {\footnotesize (-0,0189)} & {\footnotesize (-0,0188)} & {\footnotesize (-0,0118)}\\ \cmidrule{2-5}
\multirow{2}{*}{$\beta_8$} & \multirow{2}{*}{tax} & 0,00017 & 0,00017 & 0,00009\\
&  & {\footnotesize (0,00001)} & {\footnotesize (0,00001)} & {\footnotesize (0,00001)}\\ \cmidrule{2-5}
\multirow{2}{*}{$\beta_9$} & \multirow{2}{*}{ptratio} & -0,01651 & -0,01618 & -0,00862\\
&  & {\footnotesize (-0,0144)} & {\footnotesize (-0,0140)} & {\footnotesize (-0,0058)}\\ \cmidrule{2-5}
\multirow{2}{*}{$\beta_{10}$} & \multirow{2}{*}{black} & -0,00017 & -0,00017 & -0,00014\\
&  & {\footnotesize (0,00002)} & {\footnotesize (-0,00002)} & {\footnotesize (0,00001)}\\ \cmidrule{2-5}
\multirow{2}{*}{$\beta_{11}$} & \multirow{2}{*}{medv} & -0,04293 & -0,04274 & -0,04115\\
&  & {\footnotesize (-0,0456)} & {\footnotesize (-0,0455)} & {\footnotesize (-0,0435)}\\ \cmidrule{2-5}
TMSE &  & 5,41396 & 5,34753 & 3,87807\\
			\hline
			\multicolumn{5}{l}{{\footnotesize  $t-values$ in parethesis are calculated by bootstrapping.}}\\
		\end{tabular}
\end{table}

\noindent Given that we are using a logit link function, a positive value of $\tilde{\beta}_i$ indicates a positive effect of the explanatory variable $x_i$ and an increase in $\mu$. \\

The table also presents the TMSE of the estimators, calculated by bootstrap. LASSO has the smallest TMSE, followed by RR.

\section{Some Final Remarks }

In this article we have developed the RR and LASSO estimators in the context of the beta regression model, using a logit link function. The approach taken here is that of a penalized likelihood function to estimate the mean function $\mu$ and the precision parameter $\phi$ of the beta distribution. We started by using the  Newton Raphson method to obtain the maximum likelihood estimators of $\beta$ and $\phi$. Subsequently, taking the maximum likelihood estimator ($\hat{\phi}$), as the estimator of $\phi$, the RR ($\hat{\beta}_{ridge}$) and the LASSO ($\hat{\beta}_{lasso}$) estimators were found. To obtain the LASSO use is made of the coordinate descent method.

Regarding the shrinkage or regularization parameter of the RR estimator we tried several methods to select it. We found that the proposal of Abonazel \& Taha \cite{taha}:  $k=k_{med}$ performed best from the TMSE point of view. For LASSO, we used cross validation.

A number of simulation experiments were carried out to evaluate the performance of the proposed estimators. We found that RR and LASSO are useful alternatives to the maximum likelihood estimator. Overall RR seems to perform best with increasing collinearity and larger sample sizes. LASSO however,  has the advantage of selecting variables, something which is specially beneficial when some coefficients are truly zero.

Finally, the usefulness of the proposed estimators is also demonstrated through an application to a real life problem. 
Concretely, we estimated a beta regression model relating the percentage of lower status of the population in Boston to a number of economic, demographic and environmental variables, showing the best fit with LASSO.

\subsection*{Acknowledgements}
 MGN was partially supported by Fondecyt Iniciación 11200500.

\section{Appendices}

\subsection{Detailed results for ridge estimator}\label{secA1}

A first order linear approximation of the score function and applying Newton Raphson we have:
$$
\begin{array}{rcl}
	U^{k}(\hat{\beta}_{k}) &=& U^{k}(\hat{\beta}_0)+(\hat{\beta}_{k}-\hat{\beta}_0)^\prime \Omega^{k}(\hat{\beta}_0)+0(\|\hat{\beta}_{k}-\hat{\beta}_0\|).
\end{array}
$$
where $U^{k}(\beta)=U(\beta)-2k\beta$ is the score function and $\Omega^{k}(\beta)=\Omega(\beta)-2k I$ is the second derivative of $l(\beta)$. Then, equating $U^{k}(\beta_{k})$ to 0, we have
$$
\begin{array}{rcl}
	0&=&U(\hat{\beta}_0)-2k\hat{\beta}_0^{t}+(\hat{\beta}_{k}-\hat{\beta}_0)^{t}(\Omega(\hat{\beta}_0)-2k I),\\
	\hat{\beta}_{k}&=&\hat{\beta}_0-\{\Omega(\hat{\beta}_0)-2k I\}^{-1}\{U(\hat{\beta}_0)-2k\hat{\beta}_0\}\\
	&=&\{\Omega(\hat{\beta}_0)-2k I\}^{-1}\{\Omega(\hat{\beta}_0)\hat{\beta}_0 - 2k\hat{\beta}_0-U(\hat{\beta}_0)+2k\hat{\beta}_0\}\\
	&=&\{\Omega(\hat{\beta}_0)-2k I\}^{-1}\{-U(\hat{\beta}_0)+\Omega(\hat{\beta}_0)(\hat{\beta}+\Omega^{-1}(\hat{\beta}_0)U(\hat{\beta}_0))\}\\
	&=& \{\Omega(\hat{\beta}_0)-2k I\}^{-1}\{\Omega(\hat{\beta}_0)\hat{\beta}\}.
\end{array}
$$
To get our estimator we approximate $\Omega(\hat{\beta}_0)$ with the Fisher information matrix: $\mathfrak{J}_{\beta\beta} = E(-\Omega(\hat{\beta}_0)) = E\left(-\dfrac{\partial^2 l(\beta,\phi)}{\partial\beta\partial\beta^{t}}\right)=\phi X^tWX$. Hence:
\begin{eqnarray}\label{gen}
	\hat{\beta}_{k} &=& \left(X^{t} WX + 2\frac{k}{\phi}I\right)^{-1}X^{t} WX\hat{\beta}.
\end{eqnarray}
If we now set  $k^*=\dfrac{2 k}{\phi}$, we obtain the RR estimator in the beta regression model:
$$	\hat{\beta}_{ridge} =\left(X^{t}WX+k^*I\right)^{-1}X^{t}WX\hat{\beta}$$

\subsection{Detailed results for lasso estimator}\label{secA2}
Deriving first $l(\beta)$ with respect to $\beta_i$ we have:
\begin{equation}\label{log_ver}
	\dfrac{\partial l(\beta,\phi) }{\partial \beta_i} = \sum\limits_{t=1}^n \dfrac{\partial l_t(\mu_t,\phi)}{\partial \mu_t} \dfrac{\partial \mu_t}{\partial \eta_t} \dfrac{\partial \eta_t}{\partial \beta_i},
\end{equation}
\noindent where
$$
\begin{array}{rcl}
	\dfrac{\partial l_t(\mu_t,\phi)}{\partial \mu_t}&=&-\phi \psi(\mu_t\phi)+\phi \psi((1-\mu_t)\phi)+\phi \log y_t-\phi \log(1-y_t)\\
	&=&\phi(y_t^\ast-\mu_t^\ast),\\
\end{array}
$$
with $y_t^\ast=\log\left(\dfrac{y_t}{1-y_t}\right)$ and $\mu_t^\ast=\psi(\mu_t \phi)-\psi\left((1-\mu_t)\phi\right)$.
Then, denoting $g(\mu_t)=x_t^\prime \beta=\eta_t$
$$
\begin{array}{rcl}
	\dfrac{\partial l(\beta,\phi)}{\partial\beta_i}= 
	\dfrac{\partial l(\beta)}{\partial \beta} 
	=\sum\limits_{t=1}^n\phi(y_t^\ast-\mu_t^\ast)\dfrac{1}{g^\prime(\mu_t)}x_{ti}.
\end{array}
$$ 
Hence, the score function is
\begin{equation}\label{score1}
	U_\beta(\beta,\phi)=\frac{\partial l(\beta, \phi)}{\partial \beta}= \phi X^{t} T (y^\ast-\mu^\ast),
\end{equation}
\noindent where $y^\ast$ and $\mu^\ast$ are vectors gives by $y^\ast=(y_1^\ast \ldots y_n^\ast)^{t}$,  $\mu^\ast=(\mu_1^\ast \ldots \mu_n^\ast)^{t}$ and $T = \Diag\left(\frac{1}{g^\prime(\mu_1)}\ldots \frac{1}{g^\prime(\mu_n)}\right)$.
Similarly, we obtain
\begin{equation}\label{score2}
	U_\phi(\beta,\phi)=\sum \limits_{t=1}^n \psi (\phi)-\psi((1-\mu_t)\phi)+\mu_t (y_t^\ast-\mu_t^\ast)+\log (1-y_t),
\end{equation}
\noindent and finally the complete score function is $U=\left(U_\beta(\beta, \phi),U_\phi(\beta, \phi)\right)^{t}$.
The next step is to obtain the Fisher information matrix by taking the second derivative of  $l(\beta,\phi)$ with respect to $\beta$ and $\phi$ and taking expectations we have:
$$
\begin{array}{rcl}
	E\left(\dfrac{\partial^2l(\beta,\phi)}{\partial\beta_j\beta_i}\right)&=&E\left(\sum\limits_{t=1}^n \dfrac{\partial}{\partial \beta_j}\left(\dfrac{\partial l_t(\mu_t,\phi)}{\partial \mu_t}\dfrac{\partial\mu_t}{\partial\eta_t}\dfrac{\partial\eta_t}{\partial\beta_i}\right)\right),\\
	&=&E\left(\sum\limits_{t=1}^n \left(\dfrac{\partial^2 l_t(\mu_t,\phi)}{\partial^2 \mu_t}\dfrac{\partial\mu_t}{\partial\eta_t}+\dfrac{\partial l_t(\mu_t,\phi)}{\partial \mu_t}\dfrac{\partial^2\mu_t}{\partial\mu_t\eta_t}\right)\dfrac{\partial\mu_t}{\partial\eta_t}x_{ti}x_{tj}\right),\\
	&=&\sum\limits_{t=1}^n E\left(\dfrac{\partial^2 l_t(\mu_t,\phi)}{\partial^2 \mu_t}\right)\left(\dfrac{\partial\mu_t}{\partial\eta_t}\right)^2x_{ti}x_{tj}+E\left(\dfrac{\partial l_t(\mu_t,\phi)}{\partial \mu_t}\right)\dfrac{\partial^2\mu_t}{\partial\mu_t\eta_t}\dfrac{\partial\mu_t}{\partial\eta_t}x_{ti}x_{tj}.
\end{array}
$$
For properties of the beta distribution, $E\left(\dfrac{\partial l_t (\mu_t,\phi)}{\partial\mu_t}\right)=0$, eliminating the second term. Also, taking the second derivative with respect to $\mu_t$, we obtain $\dfrac{\partial^2 l_t(\mu_t,\phi)}{\partial^2 \mu_t}
	=-\phi^2\{\psi^\prime(\mu_t\phi)+\psi^\prime((1-\mu_t)\phi)\}\label{dif2_mu}$. Then
\begin{equation}\label{fisherbeta}
	E\left(\dfrac{\partial^2 l(\beta,\phi)}{\partial\beta\partial\beta^{t}}\right)=-\phi X^{t} W X,
\end{equation}
\noindent where $W=\Diag(w_1,\ldots,w_n)$ and $w_t=\phi\left\{\psi^\prime(\mu_t\phi)+\psi^\prime((1-\mu_t)\phi)\right\} \dfrac{1}{(g^\prime(\mu_t))^2}$.
Let us now consider the derivative of (\ref{score1}) with respect to $\phi$
$$
\begin{array}{rcl}
	\dfrac{\partial^2 l(\beta,\phi)}{\partial\phi \partial\beta_i}&=&\sum\limits_{t=1}^n\left[(y_t^\ast-\mu_t^\ast)-\phi \dfrac{\partial\mu_t^\ast}{\partial\phi}\right]\dfrac{1}{g^\prime(\mu_t)}x_{ti}, \\
	E\left(\dfrac{\partial^2 l(\beta,\phi)}{\partial\phi\partial\beta_i}\right)&=&\sum\limits_{t=1}^n\left[E\left[y_t^\ast-\mu_t^\ast\right]-\phi \dfrac{\partial\mu_t^\ast}{\partial\phi}\right]\dfrac{1}{g^\prime(\mu_t)}x_{ti}=-\sum c_t \dfrac{1}{g^\prime(\mu_t)}x_{ti},
\end{array}
$$
\noindent where $c_t=\phi\left[\psi^\prime(\mu_t,\phi)\mu_t-\psi^\prime\left((1-\mu_t)\phi\right)(1-\mu_t)\right]$. In matrix terms 
\begin{equation}\label{fisherphibeta}
	E\left(\dfrac{\partial^2 l(\beta,\phi)}{\partial\beta\partial\phi}\right)= -X^{t} Tc,
\end{equation}
\noindent with $c=(c_1, \ldots, c_n)^{t}$.
Finally, taking the second derivative of (\ref{score2}) with respect to $\phi$:
\begin{eqnarray}
	\dfrac{\partial^2 l(\beta,\phi)}{\partial^2\phi} & = & \sum\limits_{t=1}^n-\mu_t^2\psi^\prime(\mu_t\phi)-\psi^\prime((1-\mu_t)\phi)((1-\mu_t)^2)+\psi^\prime(\phi),\noindent\\
    E\left(\dfrac{\partial^2 l(\beta,\phi)}{\partial^2\phi}\right) & = & -\Tr(D).=-\sum\limits_{t=1}^n d_t.\label{dif2_phi}
\end{eqnarray}
where $D=\Diag(d_1,...,d_n)) , \quad d_t=\mu_t^2\psi^\prime(\mu_t\phi)+\psi^\prime((1-\mu_t)\phi)(1-\mu_t)^2-\psi^\prime(\phi)$.
From equations (\ref{log_mod}) and (\ref{logmod1}) we have $l^\prime(\beta) = \phi X^{t} T (y^\ast-\mu^\ast)$ and $l^{\prime\prime}(\beta) = -\phi X^{t} W X$, so, by applying some operations, we obtain
\begin{equation}\label{score_xb}
	l^{\prime}(\eta) = \phi T (y^\ast-\mu^\ast), \quad \mbox{ and } \quad l^{\prime\prime}(\eta)=-\phi W.
\end{equation}
Hence  a second order approximation of the log-likelihood around $\tilde{\beta}$ is
$$
\begin{array}{rcl}
	l(\beta) & \simeq & l(\tilde{\beta})+(\beta-\tilde{\beta})^{t} l^\prime(\tilde{\beta})+\frac{(\beta-\tilde{\beta})^{t} l^{\prime\prime}(\tilde{\beta})(\beta-\tilde{\beta})}{2},\\
	&\simeq& l(\tilde{\beta})+(\eta-\tilde{\eta})^{t} l^\prime(\tilde{\eta})+\dfrac{(\eta-\tilde{\eta})^{t} l^{\prime\prime}(\tilde{\eta}) (\eta-\tilde{\eta})}{2},\\
	&\simeq& l(\tilde{\beta})+\frac{1}{2}\left[\tilde{\eta}^{t} l^{\prime\prime}(\tilde{\eta}) \tilde{\eta} - \tilde{\eta}^{t} l^\prime(\tilde{\eta}) - \tilde{\eta}^{t} l^{\prime\prime}(\tilde{\eta}) \eta - l^\prime(\tilde{\eta})^{t}\tilde{\eta} 
	\right.,\\
	&& +{l^\prime(\tilde{\eta})}^{t}l^{\prime\prime}(\tilde{\eta})^{-1}l^\prime(\tilde{\eta}) + l^\prime(\tilde{\eta})^{t}(\eta) - (\eta)^{t} l^{\prime\prime}(\tilde{\eta}) (\tilde{\eta}) + (\eta)^{t} l^\prime(\tilde{\eta}), \\
	&&\left. + (\eta)^{t} l^{\prime\prime}(\tilde{\eta}) (\eta) \right] -\frac{1}{2} {l^\prime(\tilde{\eta})}^{t}l^{\prime\prime}(\tilde{\eta})^{-1}l^\prime(\tilde{\eta}),\\
	&\simeq& l(\tilde{\beta})+\frac{1}{2}\left[\tilde{\eta} -l^{\prime\prime}(\tilde{\eta})^{-1}l^{\prime}(\tilde{\eta}) - \eta \right]^{t} l^{\prime\prime}(\tilde{\eta})\left[\tilde{\eta} -l^{\prime\prime}(\tilde{\eta})^{-1}l^{\prime}(\tilde{\eta}) - \eta \right],\\
	&& -\frac{1}{2} {l^\prime(\tilde{\eta})}^{t}l^{\prime\prime}(\tilde{\eta})^{-1}l^\prime(\tilde{\eta})
\end{array}
$$
then we arrive to equation \eqref{taylor}.

\bibliographystyle{plain} 

\bibliography{biblio}

\end{document}